\documentclass[]{raa}            
\usepackage{graphicx,times}
\usepackage{natbib}

\providecommand{\bm}[1]{\mbox{\boldmath$#1$\unboldmath}}

\begin{document}

   \title{Influence of scalar-isovector $\delta$-meson field on quark phase structure in neutron stars
}

 \volnopage{ {\bf 2010} Vol.\ {\bf XX} No. {\bf XX}, 000--000}
   \setcounter{page}{1}

   \author{G.B. Alaverdyan }

   \institute{Yerevan State University, Yerevan 0025, Armenia; {\it galaverdyan@ysu.am}\\
\vs \no
   {\small Received [2010] [month] [day]; accepted [year] [month] [day] }
}

\abstract{ The deconfinement phase transition from hadronic matter
to quark matter in the interior of compact stars is investigated.
The hadronic phase is described in the framework of relativistic
mean-field (RMF) theory, when also the scalar-isovector
$\delta$-meson effective field is taken into account.  The MIT bag
model for describing a quark phase is used. The changes of the
parameters of phase transition caused by the presence of
$\delta$-meson field are explored. Finally, alterations in the
integral and structural parameters of hybrid stars due to both
deconfinement phase transition and inclusion of $\delta$-meson
field are discussed.  }

   \authorrunning{G. B. Alaverdyan }            
   \titlerunning{Influence of Scalar-Isovector $\delta$-Meson Field on Quark Phase Structure in Neutron Stars}  
   \maketitle


%
%
\section{Introduction}           
\label{sect:intro}

The structure of compact stars functionally depends on the
equation of state (EOS) of matter in a sufficiently wide range of
densities - from $7.9$ g/cm$^3$ (the endpoint of thermonuclear
burning) to one order of magnitude higher than nuclear saturation
density. Therefore, the study of properties and composition of the
matter constituents at extremely high density region is of a great
interest in both nuclear and neutron star physics. The
relativistic mean-field (RMF) theory (\citealt {Wal74,SW86,SW97})
has been effectively applied to describe the structure of finite
nuclei (\citealt{Lal97,Typel99}), the features of heavy-ion
collisions (\citealt{Ko_Li96,PR_Lal07}), and the equation of state
(EOS) of nuclear matter (\citealt{Mill95}). Inclusion of the
scalar-isovector $\delta$-meson in this theoretical scheme and
investigation of its influence on low density asymmetric nuclear
matter was realized in Refs. \cite{Kubis97,Liu02,Greco03}. At
sufficiently high density, different exotic degrees of freedom,
such as pion and kaon condensates, also deconfined quarks, may
appear in the strongly interacting matter. The modern concept of
hadron-quark phase transition is based on the feature of that
transition, that is the presence of two conserved quantities in
this transition: baryon number and electric charge
(\citealt{Gl92}). It is known that, depending on the value of
surface tension, $\sigma_{s}$, the phase transition of nuclear
matter into quark matter can occur in two scenarios
(\citealt{Heis93, Heis99}): ordinary first order phase transition
with a density jump (Maxwell construction), or formation of a
mixed hadron-quark matter with a continuous variation of pressure
and density (Glendenning construction) (\citealt{Gl92}).
Uncertainty of the surface tension values does not allow to
determine the phase transition scenario, taking place in reality.
In our recent paper (\citealt{Al09a}) in the assumption that the
transition to quark matter is a usual first-order phase
transition, described by Maxwell construction, we have shown that
the presence of the $\delta$-meson field leads to the decrease of
transition pressure $P_{0}$, of baryon number densities $n_{N}$
and $n_{Q}$.

In this article we investigate the hadron-quark phase transition
of neutron star matter, when the transition proceeds through a
mixed phase. The calculations results of the mixed phase structure
(Glendenning construction) are compared with the results of usual
first-order phase transition (Maxwell construction). Also
influence of $\delta$-meson field on phase transition
characteristics is discussed. Finally, using the EOS obtained, we
calculate the integral and structural characteristics of neutron
stars with quark degrees of freedom.


\section{Neutron star matter equation of state}
\label{sect:EOS}
\subsection{Nuclear Matter}

In this section we consider the EOS of matter in the region of
nuclear and supranuclear density ($n\geq 0.1$ fm$^{-3}$). For the
lower density region, corresponding to the outer and inner crust
of the star,  we have used the EOS of Baym-Bethe-Pethick (BBP)
(\citealt{BBP71}). To describe the hadronic phase we use the
relativistic nonlinear Lagrangian density of many-particle system
consisted of nucleons, $p$, $n$, electrons and isoscalar-scalar
($\sigma$), isoscalar-vector ($\omega$), isovector-scalar
($\delta$), and isovector-vector ($\rho$) - exchanged
mesons:\footnote{We use the natural system of units with
$\hbar=c=1$.}
\begin{eqnarray}
\label{eq1} { \cal L}=\overline{\psi}_{N}[\gamma ^{\mu }(
i\partial _{\mu }-g_{\omega }\omega _{\mu }(x)-\frac{1}{2}g_{\rho }%
\overrightarrow{\tau }_{N}\overrightarrow{\rho }_{\mu }(x)) -(
m_{N}-g_{\sigma }\sigma (x)-g_{\delta }\overrightarrow{\tau
}_{N}\overrightarrow{\delta }(x))] \psi _{N} \nonumber ~~\\
+\frac{1}{2}( \partial_{\mu }\sigma (x)\partial ^{\mu }\sigma
(x)-m_{\sigma }\sigma (x)^{2}) -U(\sigma(x)) +\frac{1}{2}m_{\omega
}^{2}\omega ^{\mu }(x)\omega _{\mu
}(x)-\frac{1}{4}\Omega _{\mu \nu}(x)\Omega ^{\mu \nu }(x)\nonumber ~~\\%
+\frac{1}{2}( \partial _{\mu }\overrightarrow{\delta }(x)\partial
^{\mu }\overrightarrow{\delta }(x)-m_{\delta }^{2}\overrightarrow{%
\delta }(x)^{2})+\frac{1}{2}m_{\rho }^{2}\overrightarrow{\rho }^{\mu }(x)%
\overrightarrow{\rho }_{\mu }(x)-\frac{1}{4}\Re_{\mu \nu
}(x)\Re^{\mu \nu}(x)\nonumber~~\\+\overline{\psi} _{e}( {i\gamma
^{\mu }\partial _{\mu}  - m_{e}})\,\psi _{e},~~
\end{eqnarray}

\noindent where $x=x_{\mu}=(t,x,y,z)$, $\sigma(x)$, $\omega _{\mu
}(x)$, $\overrightarrow{\delta }(x)$, and $\overrightarrow{\rho
}^{\mu }(x)$ are the fields of the $\sigma$, $\omega$, $\delta$,
and $\rho$ exchange mesons, respectively, $U(\sigma)$ is the
nonlinear part of the potential of the $\sigma$-field, given by
\citealt{Bog77}
\begin{equation}
\label{eq2}
U(\sigma)=\frac{b}{3}~m_{N}(g_{\sigma}\sigma)^{3}+\frac{c}{4}~(g_{\sigma}\sigma)^{4},
\end{equation}

\noindent $m_{N}$, $m_{e}$, $m_{\sigma}$, $m_{\omega}$,
$m_{\delta}$, $m_{\rho}$ are the masses of the free particles,
$\psi _{N} = \left({{\begin{array}{*{20}c} {\psi _{p}}  \hfill \\
{\psi _{n}} \hfill \\\end{array}} } \right)$ is the isospin
doublet for nucleonic bispinors, and $\overrightarrow{\tau}$ are
the isospin $2\times2$ Pauli matrices. The symbol
"$\overrightarrow{}$" denote vectors in isotopic spin space. This
Lagrangian also includes antisymmetric tensors of the vector
fields $\omega _{\mu }(x)$ and $\overrightarrow{\rho}_{\mu }(x)$
given by
\begin{equation}
\label{eq3} \Omega _{\mu \nu}  \left( {x} \right) = \partial
_{\mu}  \omega _{\nu} \left( {x} \right) - \partial _{\nu}  \omega
_{\mu}  \left( {x} \right),\quad \;\Re _{\mu \nu}  \left( {x}
\right) = \partial _{\mu}  \overrightarrow{\rho}_{\nu}  \left( {x}
\right) -
\partial _{\nu}  \overrightarrow{\rho}_{\mu}\left( {x} \right).
\end{equation}

\noindent In the RMF theory, the meson fields
$\sigma\left({x}\right)$, $\omega_{\mu}\left({x}\right)$,
$\overrightarrow{\delta}\left({x}\right)$ and
$\overrightarrow{\rho}_{\mu} \left({x}\right)$ are replaced by the
effective mean-fields $\overline{\sigma} $,
$\overline{\omega}_{\mu}$ , $\overline{\overrightarrow{\delta }}$
and $\overline {\overrightarrow{\rho}}_{\mu}$.

This Lagrangian density (1) contains the meson-nucleon coupling
constants, $g_{\sigma }$, $g_{\omega }$, $g_{\rho }$ and
$g_{\delta}$, also parameters of $\sigma$-field self-interacting
terms, $b$ and $c$. In our calculations we take for the $\delta$
coupling constant,
$a_{\delta}=\left(g_{\delta}/m_{\delta}\right)^2=2.5$ fm$^2$, as
in Refs. \cite{Liu02,Greco03,Al09a}, for the bare nucleon mass,
$m_{N}=938.93$ MeV, for the nucleon effective mass,
$m_{N}^{\ast}=0.78~m_{N}$, for the baryon number density at
saturation, $n_{0}=0.153$ fm$^{-3}$, for the binding energy per
baryon, $f_{0}=-16.3$ MeV, for the incompressibility modulus,
$K=300$ MeV, and for the asymmetry energy, $E_{sym}^{(0)}=32.5$
MeV. Five other constants
$a_{\sigma}=\left(g_{\sigma}/m_{\sigma}\right)^2$,
$a_{\omega}=\left(g_{\omega}/m_{\omega}\right)^2$,
$a_{\rho}=\left(g_{\rho}/m_{\rho}\right)^2$, $b$ and $c$ then can
be numerically determined (\citealt{Al09a}).

\begin{table}[t]
\small \centering
\begin{minipage}[]{120mm}
\caption[]{ Model Parameters with and without $\delta$ -Meson
Field}\label{Table 1}\end{minipage}
\tabcolsep 3mm
 \begin{tabular}{lllllll}
  \hline\noalign{\smallskip}
& $a_{\sigma }$ (\ fm$^{2}$) & $a_{\omega }$ (\ fm$^{2}$) & $a_{\delta }$ (\ fm$^{2}$) & $a_{\rho }$ (\ fm$^{2}$) & $b$ (\ fm$^{-1}$) & $c$   \\
  \hline\noalign{\smallskip}
$RMF\sigma \omega \rho \delta $ & $9.154$ & $4.828$ & $2.5$ &$13.621$ & $ 1.654\cdot 10^{-2}$ & $1.319\cdot 10^{-2}$ \\
$RMF\sigma \omega \rho $        & $9.154$ & $4.828$ & $0$   & $4.794$ & $1.654\cdot 10^{-2}$  & $1.319\cdot 10^{-2}$ \\
  \noalign{\smallskip}\hline
\end{tabular}
\end{table}

In Table 1 we list the values of the model parameters with and
without the isovector-scalar $\delta$ meson interaction channel
(The models $RMF\sigma\omega\rho\delta$ and $RMF\sigma\omega\rho$,
respectively).

The knowledge of the model parameters makes it possible to solve
the set of four equations in a self-consistent way and to
determine the re-denoted mean-fields, $\sigma \equiv
g_{\sigma}\bar {\sigma}$, $\omega \equiv g_{\omega}\bar
{\omega_{0}}$, $\delta \equiv g_{\delta}\bar{\delta}^{\small(3)}$,
and $\rho \equiv g_{\rho}\bar {\rho_{0}}^{(3)}$, depending on
baryon number density $n$ and asymmetry parameter
$\alpha=(n_n-n_p)/n$. The standard QHD procedure allows to obtain
expressions for energy density $\varepsilon(n,\alpha)$ and
pressure $P(n,\alpha)$ of nuclear $npe$ plasma:

\begin{eqnarray}{}
\varepsilon_{NM} ({n,\alpha},\mu_{e}) =
\frac{{1}}{{\pi^{2}}}\int\limits_{0}^{k_{-} (n,\alpha)} {\sqrt
{k^{2} + m_{p}^{\ast}( {\sigma,\delta})^{2}}}~ k^{2}dk
\nonumber\\+ \frac{{1}}{{\pi^{2}}}\int\limits_{0}^{k_{+}
(n,\alpha)} {\sqrt {k^{2} + m_{n}^{\ast}(
{\sigma,\delta})^{2}}}~k^{2}dk  \nonumber\\+
\frac{{b}}{{3}}\,m_{N} \,\sigma ^{3} + \frac{{c}}{{4}}\,\sigma
^{4} + \frac{{1}}{{2}}\left( {\,\frac{{\sigma ^{\,2}}}{{a_{\sigma}
} } + \frac{{\omega ^{2}}}{{a_{\omega} } } + \frac{{\delta
^{\,2}}}{{a_{\delta} } } + \frac{{\rho ^{\,2}}}{{a_{\rho} }
}}\right)\nonumber\\+\frac{{1}}{{\pi^{2}}}\int\limits_{0}^{\sqrt{\mu_{e}^2-m_{e}^2}}
{\sqrt {k^{2} + {m_{e}}^{2}}}~ k^{2}dk,
\end{eqnarray}

\begin{eqnarray}{}
 P_{NM} ({n,\alpha},\mu_{e}) = \frac{1}{\pi
^{2}}\int\limits_{0}^{k_{-} (n,\alpha)} {\left( \sqrt {k_{-}
(n,\alpha)^2 + m_{p}^{\ast}({\sigma,\delta})^{2}}- \sqrt {k^{2} + m_{p}^{\ast}( {\sigma,\delta})^{2}}\right)\;}k^{2}dk  \nonumber \\
+\frac{{1}}{{\pi ^{2}}}\int\limits_{0}^{k_{+} (n,\alpha)}
{\left( \sqrt {k_{+} (n,\alpha)^2 + m_{n}^{\ast}({\sigma,\delta})^{2}}- \sqrt {k^{2} + m_{n}^{\ast}( {\sigma,\delta})^{2}}\right)}k^{2}dk \nonumber \\
- \frac{{b}}{{3}}\,m_{N} \,\sigma ^{3} - \frac{{c}}{{4}}\,\sigma
^{4}+ \frac{{1}}{{2}}\left( { - \frac{{\sigma ^{2}}}{{a_{\sigma} }
} + \,\frac{{\omega ^{2}}}{{a_{\omega }} } - \frac{{\delta
^{2}}}{{a_{\delta} } } + \frac{{\rho
^{2}}}{{a_{\rho}}}}\right)\nonumber\\+ \frac{{1}}{{3\pi ^{2}}}\mu
_{e} \left( {\mu _{e} ^{2} - m_{e}
^{2}}\right)^{3/2}-\frac{{1}}{{\pi^{2}}}\int\limits_{0}^{\sqrt{\mu_{e}^2-m_{e}^2}}
{\sqrt {k^{2} + {m_{e}}^{2}}}~ k^{2}dk,
\end{eqnarray}

\noindent where $\mu_{e}$ is the chemical potential of electrons,
\begin{equation}
m_{p}^{\ast}({\sigma,\delta})=m_{N}-\sigma-\delta,~~~m_{n}^{\ast}({\sigma,\delta})=m_{N}-\sigma+\delta
\end{equation}
\noindent are the effective masses of the proton and neutron,
respectively, and
\begin{equation}
k_{\pm}(n,\alpha)=
\left(\frac{{3\pi^{2}n}}{{2}}(1\pm\alpha)\right)^{1/3}.
\end{equation}

The chemical potentials of the proton and neutron are given by
\begin{eqnarray}
\label{} \mu_{p}(n,\alpha)=\sqrt {k_{-} (n,\alpha)^{2} + m_{p}^{\ast}({\sigma,\delta})^{\,2}} + \omega + \frac{1}{2}\rho,\\
\mu_{n}(n,\alpha)=\sqrt {k_{+} (n,\alpha)^{2} +
m_{n}^{\ast}({\sigma,\delta})^{\,2}}+ \omega - \frac{1}{2}\rho.
\end{eqnarray}
\begin{figure}[h]
\begin{center}
  \begin{minipage}[h]{0.47\linewidth}
   \begin{center}
   \includegraphics[width=0.9\textwidth]{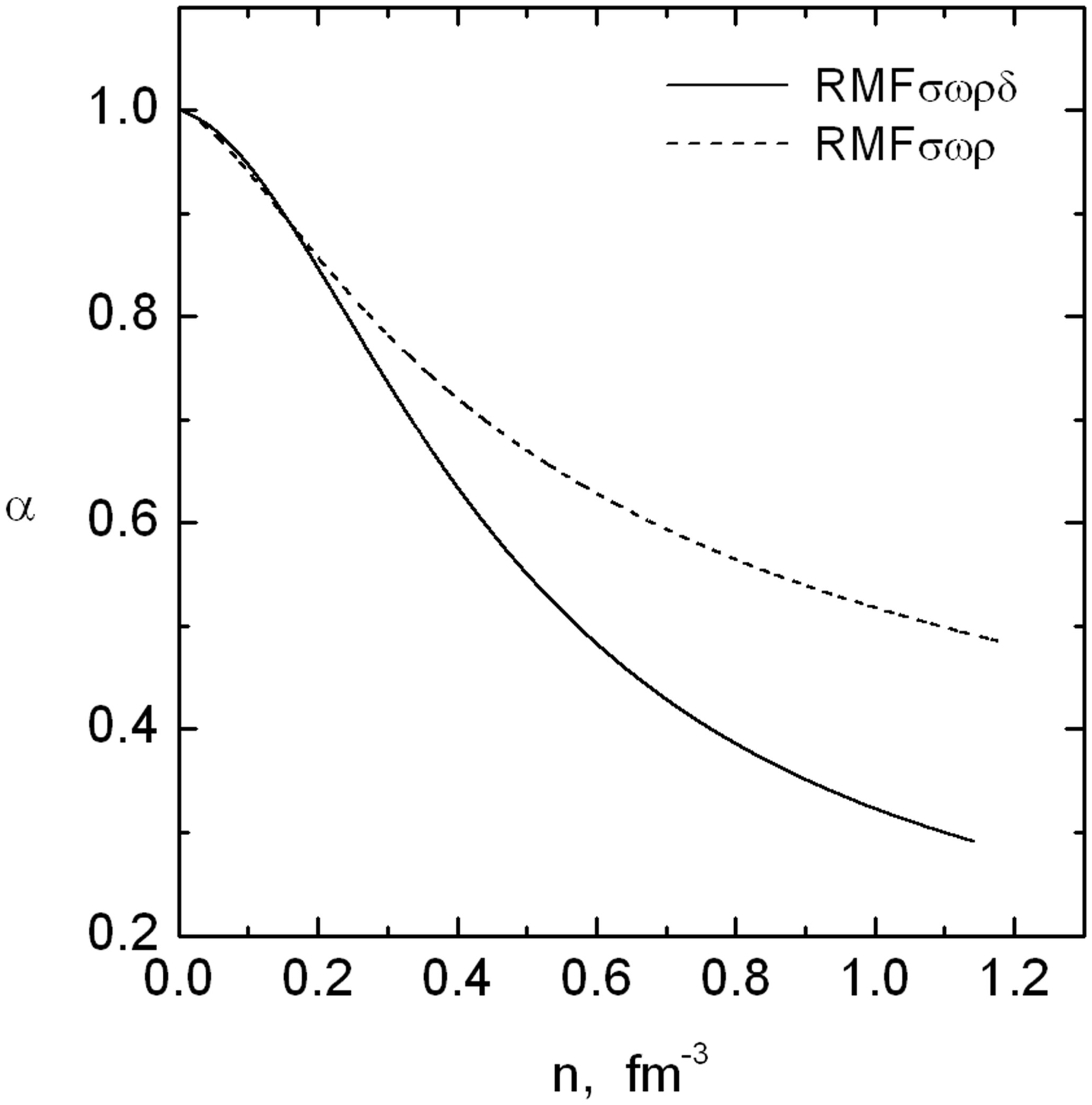}
   \caption {\small{The asymmetry parameter as a function of the
     baryon number density $n$ for a $\beta$ -equilibrium charge-neutral $npe$
    -plasma. The solid and dashed curve correspond to the $RMF\sigma \omega \rho
    \delta$ and $RMF\sigma \omega \rho$ models, respectively.}}
    \end{center}
  \end{minipage}\label{Fig1} \hfil\hfil
  \begin{minipage}[h]{0.47\linewidth}
   \begin{center}\bigskip\,
   \includegraphics[width=0.8 \textwidth]{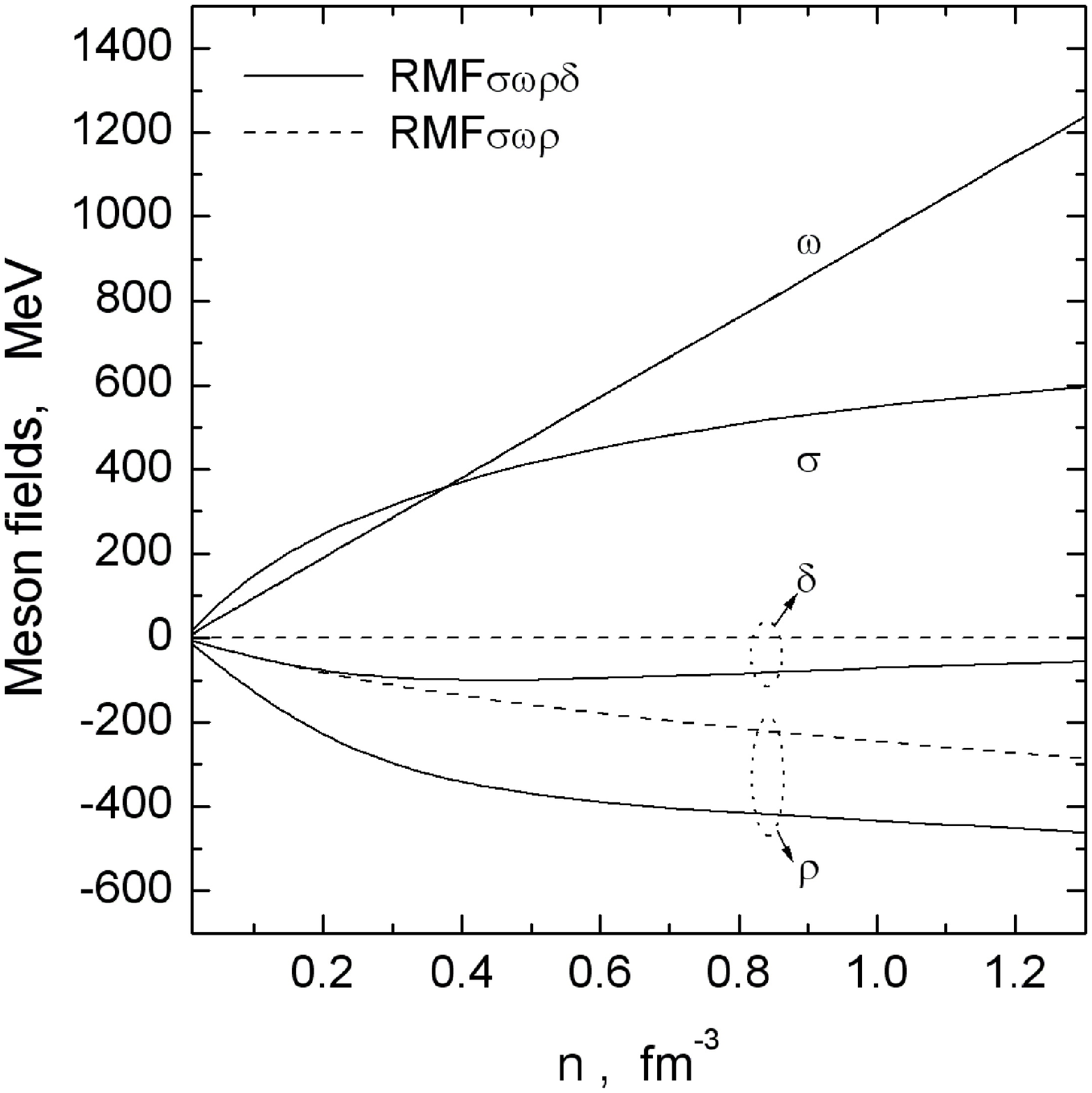}
   \caption {\small{Re-denoted meson mean-fields as a function of the baryon number
   density $n$ in case of a $\beta$-equilibrium charge-neutral $npe$-plasma with and
   without $\delta$-meson field.The solid and dashed curves correspond to the $RMF\sigma \omega \rho
    \delta$ and $RMF\sigma \omega \rho$ models, respectively.}}
    \end{center}
  \end{minipage}\label{Fig2}
 \end{center}
\end{figure}

In Fig.1 we show the asymmetry parameter $\alpha$ for the $\beta $
-equilibrium charge-neutral $npe$-plasma, as a function of the
baryon number density, $n$ (\citealt{Al09a}). The solid and dashed
curve correspond to the $RMF\sigma \omega \rho\delta$ and
$RMF\sigma \omega \rho$ models, respectively. One can see that
asymmetry parameter falls off monotonically with the increase of
baryon number density $n$. For a fixed baryon number density $n$,
the inclusion of the $\delta$-meson effective field reduces the
asymmetry parameter $\alpha$. The presence of $\delta$-field
reduces the neutron density $n_{n}$ and increases the proton
density $n_{p}$.

In Fig.2 we plotted the effective mean-fields of exchanged mesons,
$\sigma$, $\omega$, $\rho$ and $\delta $ as a function of the
baryon number density $n$ for the charge-neutral
$\beta$-equilibrium $npe$-plasma. The solid and dashed lines
correspond to the $RMF\sigma \omega \rho \delta $ and $RMF\sigma
\omega \rho $ models, respectively.

From Fig.1 and Fig.2 one can see that the inclusion of the
scalar-isovector virtual $\delta(a_{0}(980))$ meson results in
significant changes of species baryon number densities $n_{p}$ and
$n_{n}$, as well as the $\rho$ and $\delta$ meson effective
fields. This can result in changes of deconfinement phase
transition parameters and, thus, alter the structural
characteristics of neutron stars.

The results of our analysis show that the scalar - isovector
$\delta$-meson field inclusion leads to the increase of the EOS
stiffness of nuclear matter due to the splitting of proton and
neutron effective masses, and also the increase of asymmetry
energy (for details see Ref. \citealt{Al09a}).

\subsection{Quark Matter}
To describe  the quark phase an improved version of the MIT bag
model (\citealt{Chod74}) is used, in which the interactions
between $u$, $d$ and $s$ quarks inside the bag are taken into
account in the one-gluon exchange approximation
(\citealt{FarJaf84}). The quark phase consists of three quark
flavors $u$, $d$, $s$ and electrons, which are in equilibrium with
respect to weak interactions. We choose $m_{u} = 5$ MeV, $m_{d} =
7$ MeV and $m_{s} = 150$ MeV for quark masses, and
$\alpha_{s}=0.5$ for the strong interaction constant.

\subsection{Deconfinement Phase Transition Parameters}
There are two independent conserved charges in hadron-quark phase
transition: baryonic charge and electric charge. The constituents
chemical potentials of the $npe$-plasma in $\beta$-equilibrium are
expressed through two potentials, $\mu_{b}^{(NM)}$ and
$\mu_{el}^{(NM)}$, according to conserved charges, as follows
\begin{equation}\label{chempot}
\mu_{n}=\mu_{b}^{(NM)},~~~~\mu_{p}=\mu_{b}^{(NM)}-\mu_{el}^{(NM)},
~~~~\mu_{e}=\mu_{el}^{(NM)}.
\end{equation}
In this case, the pressure $P_{NM}$, energy density
$\varepsilon_{NM}$ and baryon number density $n_{NM}$, are
functions of potentials,
 $\mu_{b}^{(NM)}$ and $\mu_{el}^{(NM)}$.

The particle species chemical potentials for $udse$-plasma in
$\beta$-equilibrium  are expressed through the chemical potentials
$\mu_{b}^{(QM)}$ and $\mu_{el}^{(QM)}$ as follows
\begin{eqnarray}{}
\mu_{u}&=&\frac{1}{3}\left(\mu_{b}^{(QM)}-2~\mu_{el}^{(QM)}\right),\nonumber\\
\mu_{d}&=&\mu_{s}=\frac{1}{3}\left(\mu_{b}^{(QM)}+\mu_{el}^{(QM)}\right),\\
\mu_{e}&=&\mu_{d}-\mu_{u}=\mu_{el}^{(QM)}.\nonumber
\end{eqnarray}

In this case, the thermodynamic characteristics, pressure
$P_{QM}$, energy density $\varepsilon_{QM}$ and baryon number
density $n_{QM}$, are functions of chemical potentials
$\mu_{b}^{(QM)}$ and ~$\mu_{el}^{(QM)}$.

The mechanical and chemical equilibrium conditions (Gibbs
conditions) for mixed phase are
\begin{equation}{}
\mu_{b}^{(QM)}=\mu_{b}^{(NM)}=\mu_{b},~~~~\mu_{el}^{(QM)}=\mu_{el}^{(NM)}=\mu_{el},
\end{equation}
\begin{equation}{}
P_{QM}(\mu_{b},~\mu_{el})=P_{NM}(\mu_{b},~\mu_{el}).
\end{equation}

The volume fraction of quark phase is
\begin{equation}
\chi=V_{QM}/\left(V_{QM}+V_{NM}\right),
\end{equation}
\noindent where $V_{QM}$ and $V_{NM}$ are volumes occupied by
quark matter and nucleonic matter, respectively.

We applied the global electrical neutrality condition for mixed
quark-nucleonic matter, according to Glendenning
(\citealt{Gl92,Gl00}),
\begin{eqnarray}{}
(1-\chi)\left[n_{p}(\mu_{b},\mu_{el})-n_{e}(\mu_{el})\right]\nonumber\\+\chi\left[\frac{2}{3}~n_{u}(\mu_{b},\mu_{el})-\frac{1}{3}~n_{d}(\mu_{b},\mu_{el})-\frac{1}{3}~n_{s}(\mu_{b},\mu_{el})-n_{e}(\mu_{el})\right]=0.
\end{eqnarray}
The baryon number density in the mixed phase is determined as
\begin{eqnarray}{}
n=(1-\chi)\left[n_{p}(\mu_{b},\mu_{el})+n_{n}(\mu_{b},\mu_{el})\right]\nonumber\\+\frac{1}{3}~\chi\left[n_{u}(\mu_{b},\mu_{el})+n_{d}(\mu_{b},\mu_{el})+n_{s}(\mu_{b},\mu_{el})\right],
\end{eqnarray}
and the energy density is
\begin{eqnarray}{}
\varepsilon=(1-\chi)\left[\varepsilon_{p}(\mu_{b},\mu_{el})+\varepsilon_{n}(\mu_{b},\mu_{el})\right]\nonumber\\+\chi\left[\varepsilon_{u}(\mu_{b},\mu_{el})+\varepsilon_{d}(\mu_{b},\mu_{el})+\varepsilon_{s}(\mu_{b},\mu_{el})\right]+\varepsilon_{e}(\mu_{el}).
\end{eqnarray}

In case of $\chi=0$, the chemical potentials $\mu_{b}^{N}$ and
$\mu_{el}^{N}$, corresponding to the lower threshold of a mixed
phase, are determined by solving Eqs. (13) and (15). This allows
to find the lower boundary parameters $P_{N}$, $\varepsilon_{N}$
and $n_{N}$. Similarly, we calculate the upper boundary values of
mixed phase parameters, $P_{Q}$, $\varepsilon_{Q}$ and $n_{Q}$,
for $\chi=1$. The system of Eqs. (13), (15), (16) and (17) makes
it possible to determine EOS of mixed phase between this critical
states.
\begin{table}[t]
\small \centering
\begin{minipage}[]{120mm}
\caption[]{The Mixed Phase Threshold Parameters with and without
$\delta$ -Meson Field for Bag Parameter Values, $B=60$ MeV/fm$^3$
and $B=100$ MeV/fm$^3$}\end{minipage}
\tabcolsep 3mm
\begin{tabular}{lcccccc}
\hline\noalign{\smallskip} Model& $n_{N}$ & $n_{Q}$  & $P_{N}$ &
$P_{Q}$ & $\varepsilon_{N}$ & $\varepsilon_{Q}$ \\
 & (fm$^{-3})$ & (fm$^{-3}$) & (MeV/fm$^{3}$) & (MeV/fm$^{3}$) &
(MeV/fm$^{3}$) & (MeV/fm$^{3}$) \\
\noalign{\smallskip}\hline\noalign{\smallskip}
  B60$\sigma\omega\rho\delta$  & 0.0771  & 1.083   &  0.434   & 327.745  &  72.793   & 1280.884 \\
  B60$\sigma \omega \rho $     & 0.0717  & 1.083   &  0.336   & 327.747  &  67.728   & 1280.889 \\
  B100$\sigma\omega\rho\delta$ & 0.2409  & 1.448   & 16.911   & 474.368  & 235.029   & 1889.336 \\
  B100$\sigma \omega \rho    $ & 0.2596  & 1.436   & 18.025   & 471.310  & 253.814   & 1870.769 \\

  \noalign{\smallskip}\hline\noalign{\smallskip}
\end{tabular}
\end{table}

Note, that in the case of an ordinary first-order phase transition
both nuclear and quark matter are assumed to be separately
electrically neutral, and at some pressure $P_{0}$, corresponding
to the coexistence of the two phases, their baryon chemical
potentials are equal, i.e.,
\begin{equation}
\label{eq45} \mu _{NM} \left( {P_{0}}  \right) = \mu _{QM} \left(
{P_{0}}  \right).
\end{equation}
Such phase transition scenario is known as phase transition with
constant pressure (Maxwell construction).

Table 2 represents the parameter sets of the mixed phase with and
without $\delta$-meson field. It is shown that the presence of
$\delta$-field alters the threshold characteristics of the mixed
phase. For $B=60$ MeV/fm$^3$ the lower threshold parameters,
$n_{N}$, $\varepsilon_{N}$ and $P_{N}$, are increased, meanwhile
the upper ones, $n_{Q}$, $\varepsilon_{Q}$ and $P_{Q}$, are slowly
decreased. For $B=100$ MeV/fm$^3$ this behavior changes to
opposite.

\begin{figure}[h]
\begin{center}
  \begin{minipage}[h]{0.47\linewidth}
   \begin{center}
   \includegraphics[width=0.8\textwidth]{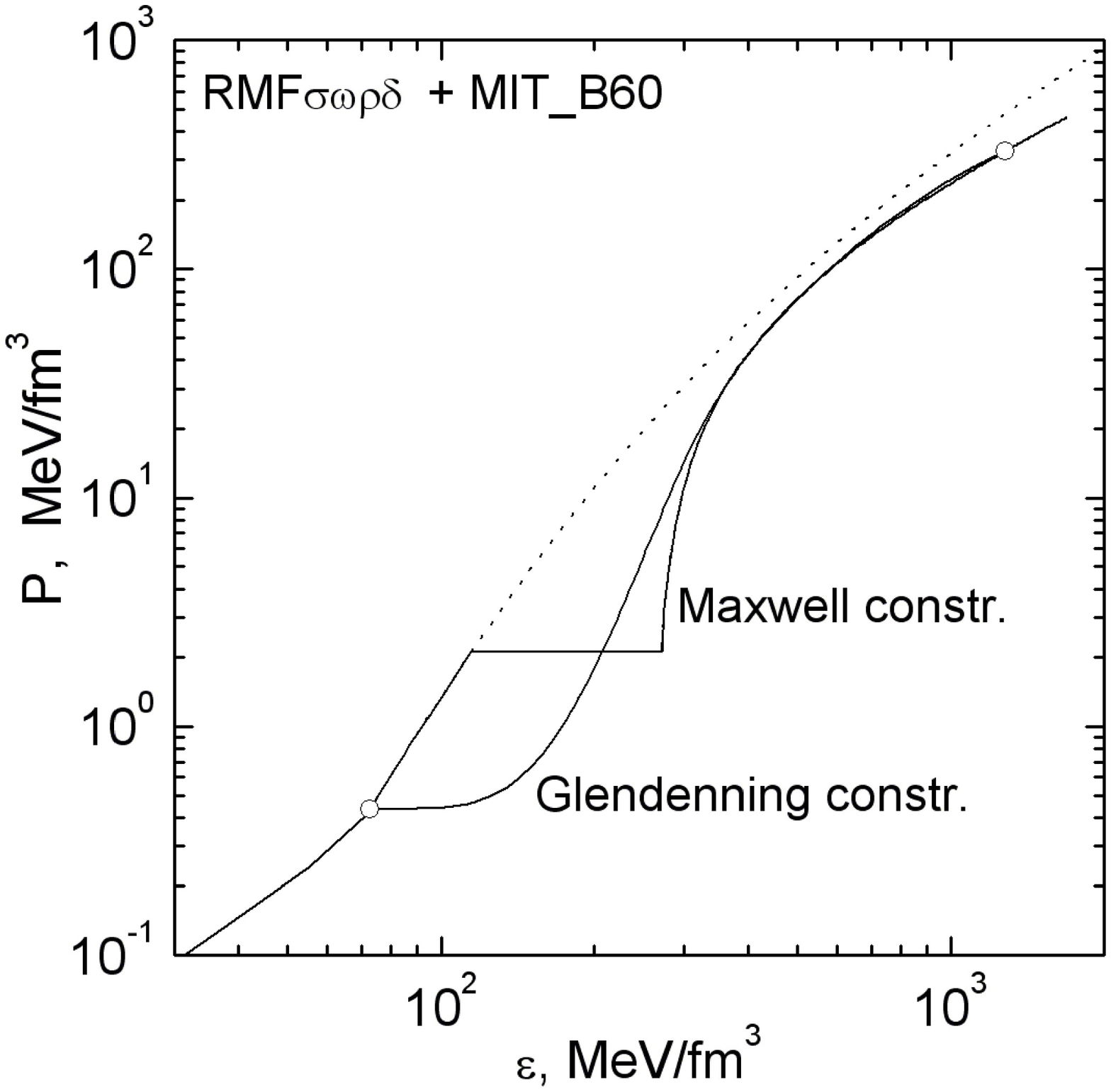}
   \caption {\small{EOS of neutron star matter with the deconfinement
   phase transition for a bag constant $B=60$ MeV/fm$^3$. For
   comparison we plot both the Glendenning and Maxwell
   constructions. Open circles represent the mixed phase boundaries.}}
    \end{center}
  \end{minipage}\label{Fig3} \hfil\hfil
  \begin{minipage}[h]{0.47\linewidth}
   \begin{center}
   \includegraphics[width=0.84\textwidth]{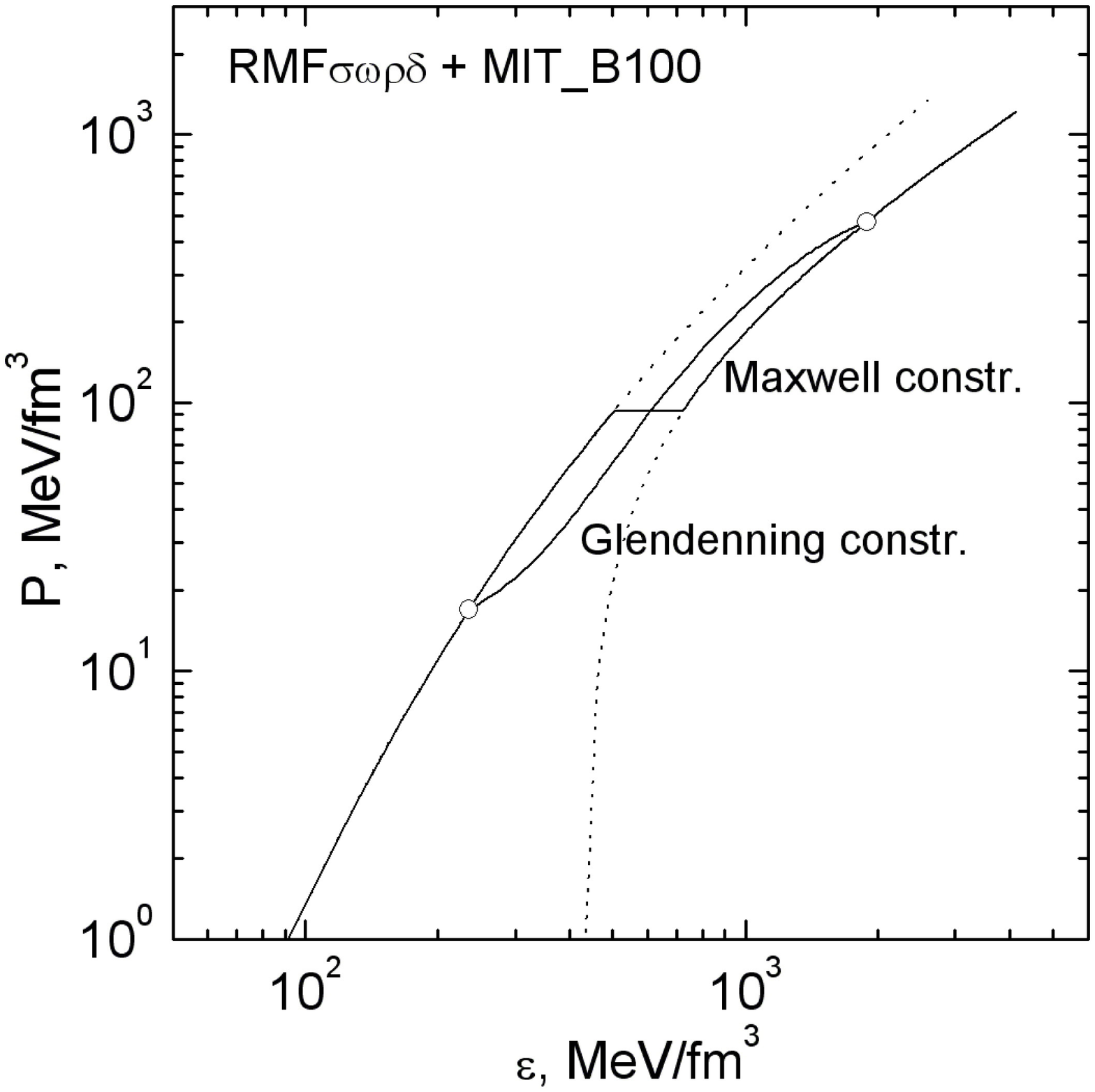}
   \caption {\small{As in Fig 3, but for $B=100$ MeV/fm$^3$.} \bigskip\bigskip\bigskip\bigskip}
    \end{center}
  \end{minipage}\label{Fig4}
 \end{center}
\end{figure}
In Fig.3 and Fig.4 we plot the EOS of compact star matter with
deconfinement phase transition for two values of bag constant,
$B=60$ MeV/fm$^3$ and $B=100$ MeV/fm$^3$, respectively. The dotted
curves correspond to pure nucleonic and quark matters without any
phase transition, while the solid lines correspond to two
alternative phase transition scenarios. Open circles show the
boundary points of the mixed phase.
\begin{figure}[h]
\begin{center}
  \begin{minipage}[h]{0.47\linewidth}
   \begin{center}
  \includegraphics[width=0.8\textwidth]{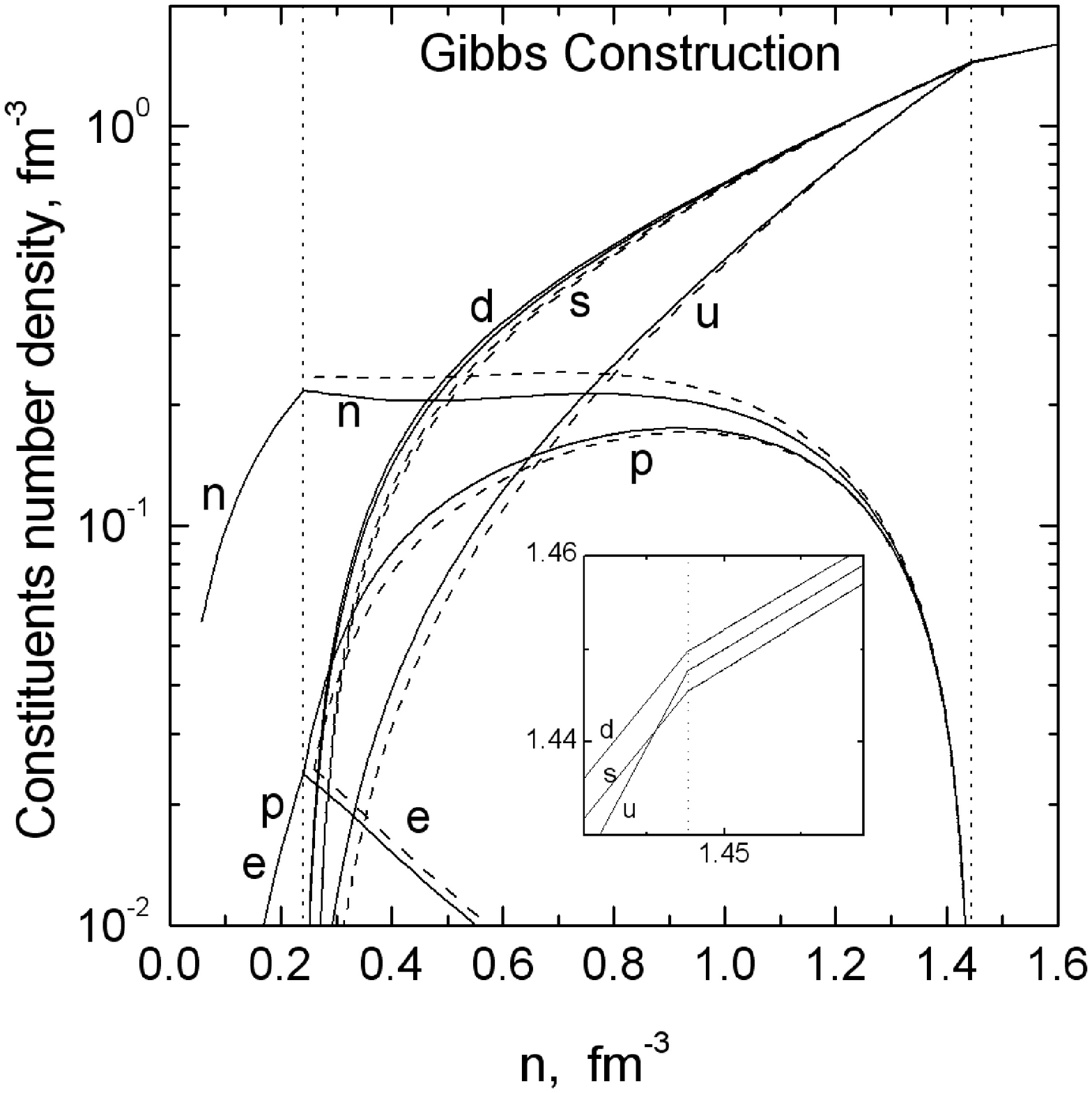}
   \caption {\small{Constituents number density versus baryon number density $n$
   for $B=100$ MeV/fm$^3$ in case of Glendenning construction. Vertical dotted lines
   represent the mixed phase boundaries. The dashed curves show appropriate results of the model without $\delta$-meson field.}}
    \end{center}
  \end{minipage}\label{Fig5} \hfil\hfil
  \begin{minipage}[h]{0.47\linewidth}
   \begin{center}
 \includegraphics[width=0.85\textwidth]{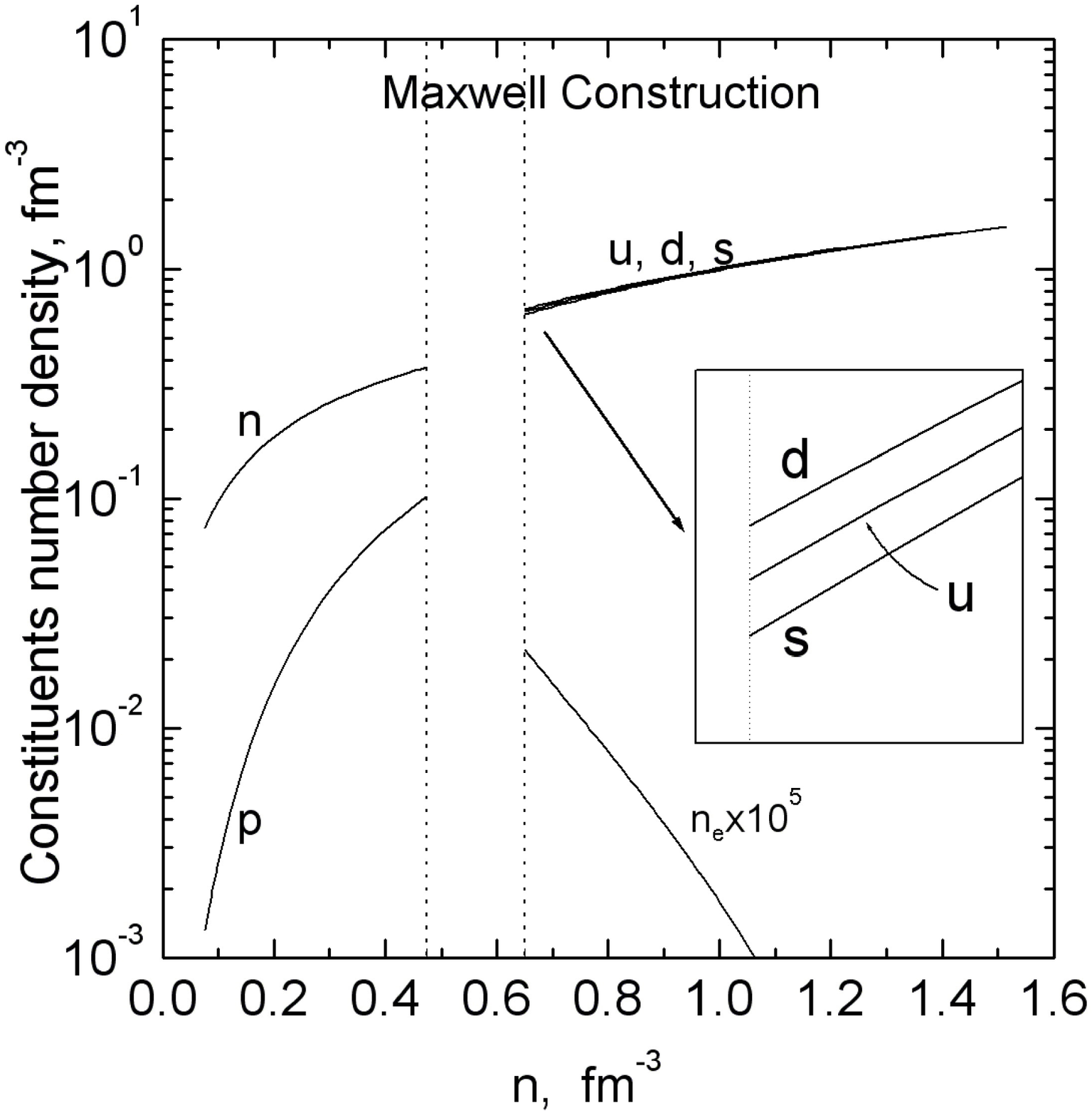}
   \caption {\small{As in Fig 5, but for Maxwell construction. Vertical dotted lines represent the density jump boundaries.
   } \bigskip\bigskip}
    \end{center}
  \end{minipage}\label{Fig6}
 \end{center}
\end{figure}

In Fig.5 we plot the particle species number densities as a
function of baryon density $n$ for Glendenning construction.
Quarks appear at the critical density $n_{N}=0.241$ fm$^{-3}$. The
hadronic matter completely disappears at $n_{Q}=1.448$ fm$^{-3}$,
where the pure quark phase occurs. The solid curves correspond to
the case, when also the $\delta$- meson effective field is taken
into account besides $\sigma,~ \omega,~ \rho$ meson fields  (model
$B100\_~\sigma\omega\rho\delta$). The dashed curves represent the
results in case when we neglect the $\delta$-meson field (model
$B100\_~\sigma\omega\rho$). One can see that inclusion of the
$\delta$-meson field leads to the increase of number densities of
quarks and protons, and simultaneously to the reduction of number
densities of neutrons and electrons. In Table 2 we have already
shown the mixed phase boundaries changes caused by the inclusion
of the $\delta$ - meson effective field.

Fig.6 shows the constituents number density as a function of
baryon number density $n$ for $B=100$ MeV/fm$^3$, when phase
transition is described according to Maxwell construction. Maxwell
construction leads to the appearance of a discontinuity. In this
case, the charge neutral nucleonic matter at baryon density
$n_{1}=0.475$ fm$^{-3}$ coexists with the charge neutral quark
matter at baryon density $n_{2}=0.650$ fm$^{-3}$. Thus, the
density range $n_{1}<n<n_{2}$ is forbidden. In case of Maxwell
construction, the chemical potential of electrons, $\mu_{e}$, has
a jump at the coexistence pressure $P_{0}$. Notice, that such
discontinuity behavior takes place only in usual first-order phase
transition, i.e., in the Maxwell construction case.

\section{Properties of hybrid stars}

Using the EOS obtained in previous section, we calculate the
integral and structural characteristics of neutron stars with
quark degrees of freedom.

The hydrostatic equilibrium properties of spherical symmetric and
isotropic compact stars in general relativity is described by the
Tolman-Oppenheimer-Volkoff(TOV) equations (\citealt{Tol39,OV39}):
\begin{eqnarray}\label{TOV}
\frac{dP}{dr}&=&-\frac{G}{r^{2}}\frac{(P+\varepsilon)(m+4\pi
r^{3}P)}{1-2G~m/r},\\
\frac{dm}{dr}&=&4\pi r^2 \varepsilon,
\end{eqnarray}
\noindent where $G$ is the gravitational constant, $r$ is the
distance from the center of star, $m(r)$ is the mass inside a
sphere of radius $r$, $P(r)$ and $\varepsilon(r)$ are the pressure
and energy density at the radius $r$, respectively. To integrate
the TOV equations, it is necessary to know the EOS of neutron star
matter in a form $\varepsilon(P)$. Using the neutron star matter
EOS, obtained in previous section, we have integrated the
Tolman-Oppenheimer Volkoff equations and obtained the
gravitational mass $M$ and the radius $R$ of compact stars (with
and without quark degrees of freedom) for the different values of
central pressure, $P_{c}$.
\begin{figure}[h]
\begin{center}
  \begin{minipage}[h]{0.47\linewidth}
   \begin{center}
   \includegraphics[width=0.97\textwidth]{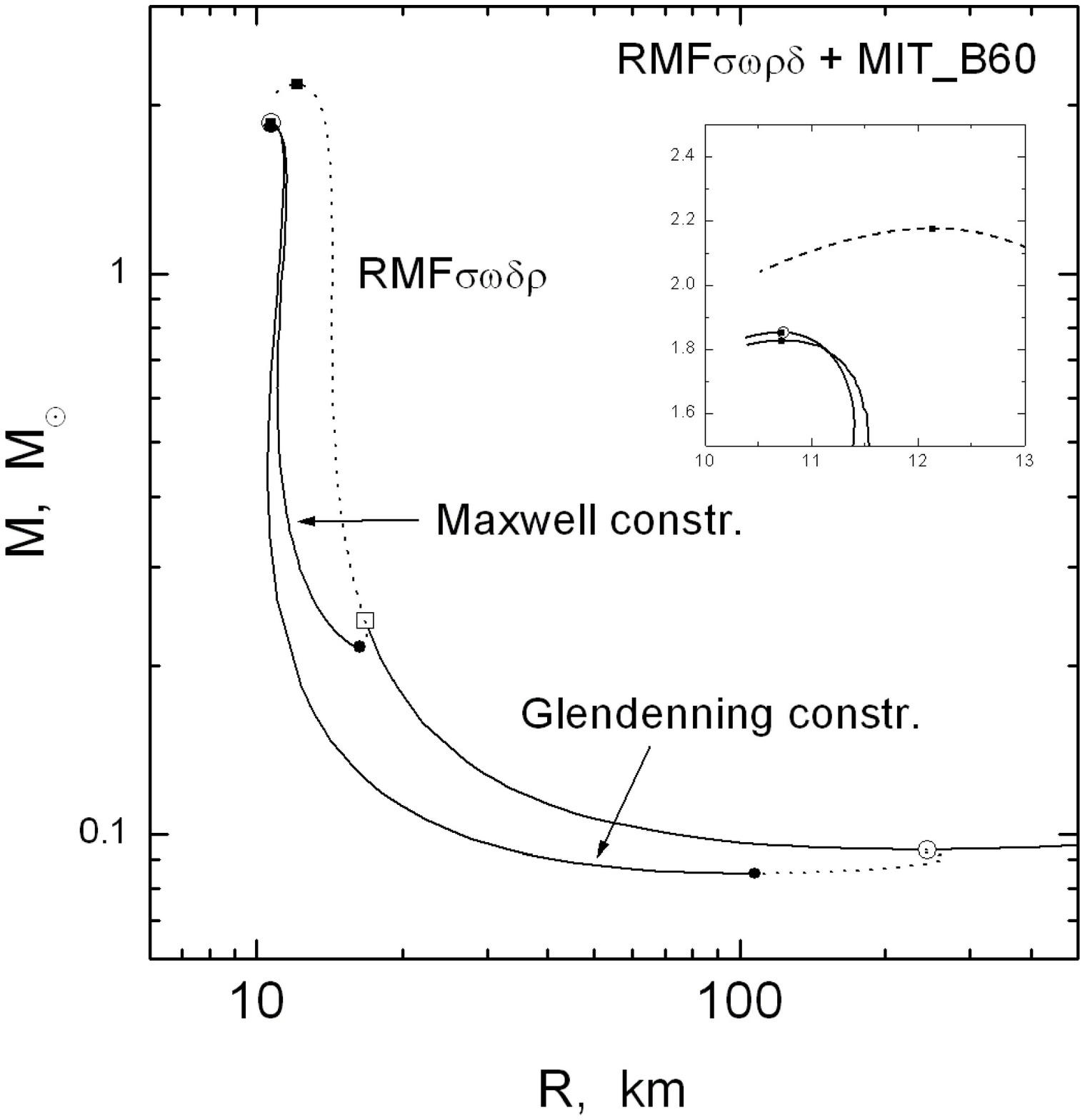}
   \caption {\small{The mass-radius relation of neutron star with different deconfinement
   phase transition scenarios for the bag constant $B=60$ MeV/fm$^3$. Open circles and squares denote
   the critical configurations for Glendenning and Maxwellian type transitions, respectively.
   Solid circles and squares denote hybrid stars with minimal and maximal masses, respectively.}}
    \end{center}
  \end{minipage}\label{Fig7} \hfil\hfil
  \begin{minipage}[h]{0.47\linewidth}
   \begin{center}
   \includegraphics[width=0.88\textwidth]{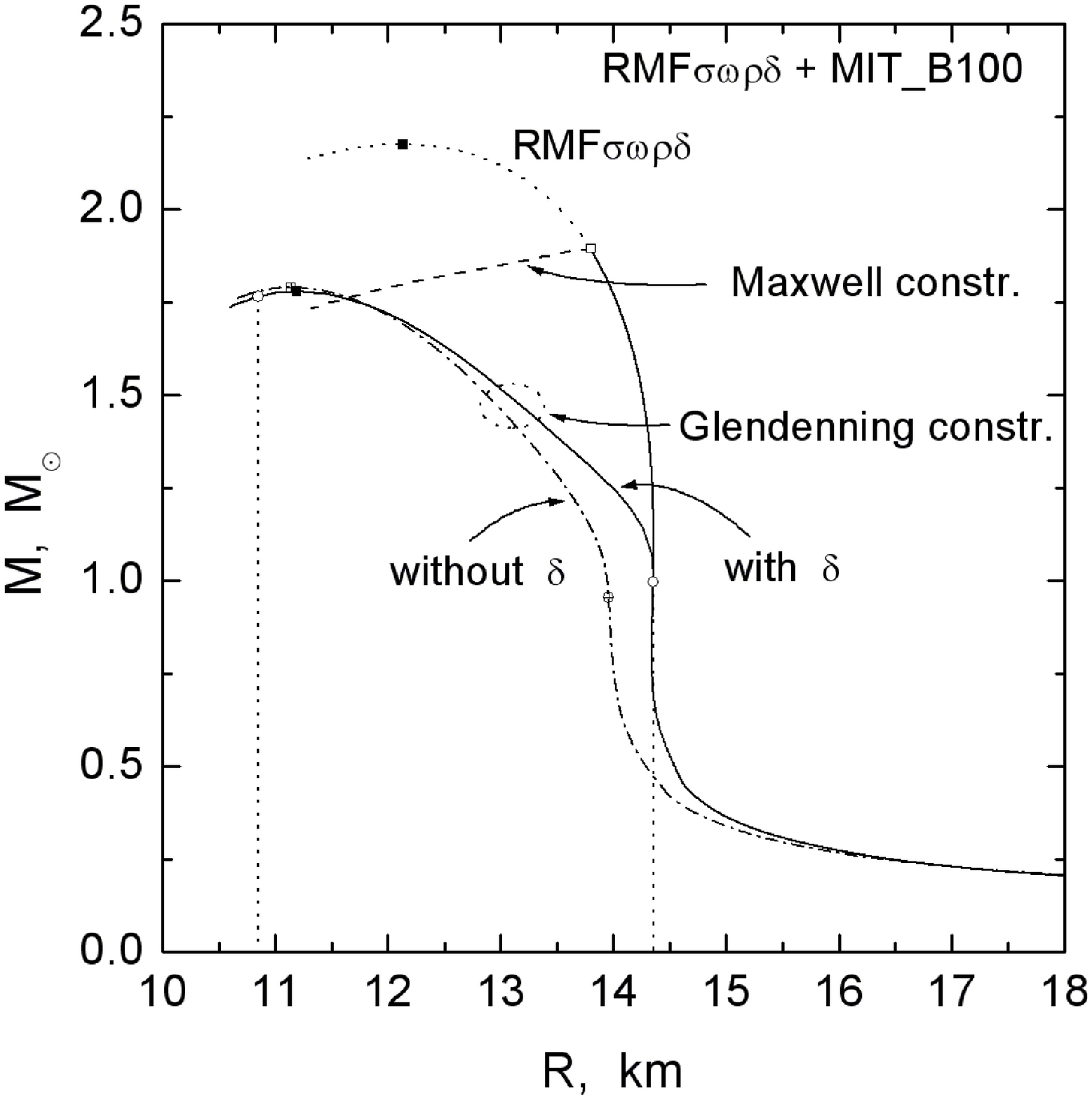}
   \bigskip\caption {\small{As in Fig 7, but for $B=100$ MeV/fm$^3$. The mass-radius relation in case
   of the Glendennig construction without the $\delta$-meson effective field is also displayed for comparison (dash-dotted curve).\bigskip}}
    \end{center}
  \end{minipage}\label{Fig8}
 \end{center}
\end{figure}

Fig.7 and Fig.8 illustrate the $M(R)$ dependence of neutron stars
for the two values of bag constant $B=60$ MeV/fm$^3$ and $B=100$
MeV/fm$^3$, respectively. We can see, that the behavior of
mass-radius dependence significantly differs for the two types of
phase transitions. Fig.7 shows, that for $B=60$ MeV/fm$^3$ there
are unstable regions, where $dM/dP_{c}<0$ between two stable
branches of compact stars, corresponding to configurations with
and without quark matter. In this case, there is a nonzero minimum
value of the quark phase core radius. Accretion of matter on a
critical neutron star configuration will then result in a
catastrophic rearrangement of the star, forming a star with a
quark matter core. The range of mass values for stars, containing
the mixed phase, is $[0.085 M_{\odot}; 1.853M_{\odot}]$ for $B=60$
MeV/fm$^3$, and is $[0.997M_{\odot}; 1.780M_{\odot}]$ for $B=100$
MeV/fm$^3$. In case of Maxwellian type phase transition, the
analogous range is $[0.216M_{\odot}; 1.828M_{\odot}]$ for $B=60$
MeV/fm$^3$. From Fig.8 one can observe, that in case of $B=100
$MeV/fm$^3$, the star configurations with deconfined quark matter
are unstable. Thus, the stable neutron star maximum mass is
$1.894M_{\odot}$. Our analysis show, that for $B=100 $MeV/fm$^3$,
the pressure upper threshold value for mixed phase is larger than
the pressure, corresponding to the maximum mass configuration.
Hence, in this case, the mixed phase can exist in the center of
compact stars, but no pure quark matter can exist. The dash-dotted
curve in Fig.8 represents the results in case when we neglect the
$\delta$-meson field (model $B100\_~\sigma\omega\rho$). One can
see that for a fixed gravitational mass the star with
$\delta$-meson field has larger radius than the corresponding star
without the $\delta$-meson field. Influence of $\delta$-meson
field on the hybrid star properties is demonstrated in Table 3,
where we display the hybrid star properties with and without
$\delta$-meson field for minimum and maximum mass configurations.
The results show that the minimum mass of hybrid stars and
corresponding radius are increased with the inclusion of the
$\delta$-meson field. Notice that influence of $\delta$-meson
field on maximum mass configuration properties is insignificant.

\begin{table}[t]
\small \centering
\begin{minipage}[]{120mm}
\caption[]{Hybrid Star Critical Configuration Properties for
$B=100$ MeV/fm$^3$ with and without $\delta$-Meson Field}
\end{minipage}
\tabcolsep 3mm
\begin{tabular}{lllllll}
\hline\noalign{\smallskip} & & & & Minimum Mass Configuration &
Maximum Mass Configuration&
\end{tabular}
\begin{tabular}{lcccccc}
\hline\noalign{\smallskip} Model & $\varepsilon_{c}$ & $M_{min}$
& $R$ & $\varepsilon_{c}$ & $M_{max}$  & $R$\\
 & (MeV/fm$^{3}$)  & ($M_{\odot}$) & (km) &  (MeV/fm$^{3}$)& ($M_{\odot}$)  & (km) \\
\noalign{\smallskip}\hline\noalign{\smallskip}
  B100$\sigma\omega\rho\delta$ & 235.029  & 0.997   & 14.354   &  1390.77 & 1.780   & 11.190 \\
  B100$\sigma \omega \rho    $ & 253.814  & 0.955   & 13.960   &  1386.03 & 1.791   & 11.139 \\
\noalign{\smallskip}\hline\noalign{\smallskip}
\end{tabular}
\end{table}

\section{Conclusions}

In this paper we have studied the deconfinement phase transition
of neutron star matter, when the nuclear matter is described in
the RMF theory with $\delta$-meson effective field. We show that
the inclusion of scalar – isovector $\delta$-meson field terms
leads to the stiff nuclear matter EOS. In a nucleonic star both
the gravitational mass and corresponding radius of the maximum
mass stable configuration increases with the inclusion of the
$\delta$ field. The presence of scalar – isovector $\delta$-meson
field alters the threshold characteristics of the mixed phase. For
$B=60$ MeV/fm$^3$, the lower threshold parameters, $n_{N}$,
$\varepsilon_{N}$, $P_{N}$, are increased, meanwhile the upper
ones, $n_{Q}$, $\varepsilon_{Q}$, $P_{Q}$, are slowly decreased.
For $B=100$ MeV/fm$^3$ this behavior changes to opposite.

In case of the bag constant value $B=100 $MeV/fm$^3$, the pressure
upper threshold value for mixed phase is larger, than the
pressure, corresponding to the maximum mass configuration. This
means that in this case, the stable compact star can possess a
mixed phase core, but the density range does not allow to possess
a pure strange quark matter core.

Stars with $\delta$-meson field have larger radius than stars of
the same gravitational mass without the $\delta$-meson field.
Alterations of the maximum mass configuration parameters caused by
the inclusion of $\delta$-meson field is insignificant.

For the bag constant value $B=60$ MeV/fm$^3$, the maximum mass
configuration has a gravitational mass $M_{max}=1.853~M_{\odot}$
with radius $R=10.71$ km, and central density
$\rho_{c}=2.322\cdot10^{15}$ g/cm$^3$. This star has a pure
strange quark matter core with radius $r_{Q}\approx0.83$ km, next
it has a nucleon-quark mixed phase layer with a thickness of
$r_{MP}\approx9.43$ km, followed by a normal nuclear matter layer
with a thickness of $r_{N}\approx0.45$ km.

\normalem
\begin{acknowledgements}
The author would like to thank  Profs. Yu.L.Vartanyan and
G.S.Hajyan for fruitful discussions on issues related to the
subject of this research.

This work was partially supported by the Ministry of Education and
Sciences of the Republic of Armenia under grant 2008-130.

\end{acknowledgements}

\label{lastpage}

\end{document}